\documentclass[11pt, a4paper]{article}

\pdfoutput=1

\usepackage{jheppub}
\usepackage{graphicx}
\usepackage{amssymb}
\usepackage{slashed}
\usepackage{amsmath}
\usepackage{url}
\usepackage[sort&compress]{natbib}
\usepackage{hyperref} 
\hypersetup{
    colorlinks=false,       
    linkcolor=red,          
    citecolor=green,        
    filecolor=magenta,      
    urlcolor=cyan           
}

\newcommand{\be}{\begin{equation}}
\newcommand{\ee}{\end{equation}}

\newcommand{\GeV}{\ \mathrm{GeV}}
\newcommand{\eref}[1]{Eq.~(\ref{#1})}
\newcommand{\sv}{\left< \sigma v \right>}
\newcommand{\ord}[1]{O\left( #1 \right)}
\newcommand{\ex}{\mathrm{ex}}

\newcommand{\rhoex}{\rho_{\rm ex}}
\newcommand{\drhoex}{\Delta \rho_{\rm ex}}

\title{Probing the Cosmological Constant and Phase Transitions with Dark Matter}

\author[a]{Daniel Chung,}
\author[a]{Andrew Long,}
\author[b]{and Lian-Tao Wang}

\affiliation[a]{Department of Physics, University of Wisconsin, Madison, WI 53706}
\affiliation[b]{Enrico Fermi Institute and Department of Physics, University of Chicago, Chicago, IL 60637}

\emailAdd{danielchung@wisc.edu}
\emailAdd{ajlong@wisc.edu}
\emailAdd{liantaow@uchicago.edu}

\abstract{ The Standard Model and its extensions predict multiple
  phase transitions in the early universe. In addition to the
  electroweak phase transition, one or several of these could occur at
  energies close to the weak scale. Such phase transitions can leave
  their imprint on the relic abundance of TeV-scale dark matter.  In
  this paper, we enumerate several physical features of a generic
  phase transition and parameterize the effect of each on the relic
  abundance.  In particular, we include among these effects the
  presence of the scalar field vacuum energy and the cosmological
  constant, which is sensitive to UV physics.  Within the context of
  the Standard Model Higgs sector, we find that the relic abundance of 
  generic TeV-scale dark matter is affected by the vacuum energy at the order of a
  fraction of a percent.  For scalar field sectors with strong first order phase 
  transitions, an order one percent apparent tuning of coupling constants may 
  allow corrections induced by the vacuum energy to be of order unity.}


\begin{document}
\maketitle


\section{Introduction}

Phase transitions (PTs) are expected to be generic in the early universe 
\cite{Kolb:1990vq}.  The high temperature environment gives rise to 
significant corrections to the vacuum structure, and symmetries which are 
broken in the universe today can be restored at earlier times \cite{Kirzhnits:1972ut, Weinberg:1974hy, Dolan:1973qd}.  The Standard Model (SM) predicts 
early universe phase transitions in both the electroweak and QCD sectors.  
Beyond the SM, it is well-known that additional degrees of freedom can modify 
the dynamics of the electroweak phase transition significantly \cite{Profumo:2007wc, Ham:2004cf, McDonald:1993ey, Espinosa:2007qk, Carena:1996wj, Kang:2004pp, Chung:2010cd, Pietroni:1992in, Menon:2004wv, Huber:2000mg, Choi:1993cv, Davies:1996qn, Huber:2006wf, Laine:1998qk, Espinosa:1993yi, Fodor:1994sj, Cline:1998hy, Ahriche:2007jp, Ahriche:2010ny}.  Furthermore, because almost all scenarios beyond the SM 
have extended symmetries, an even richer thermal history is expected in 
general (e.g., \cite{Weinberg:1974hy,Kibble:1976sj,Guth:1981uk,Guth:1979bh,Einhorn:1980ym,Kibble:1982ae,Kusenko:1996jn,Brandenberger:1996zp,Cline:1998rc,Cline:1999wi,Shu:2006mm,Cui:2007js,Murayama:2009nj}).

Most phenomenologically viable, cosmological PTs do not leave significant 
observable signals today.  The successful and precisely measured theories 
of big bang nucleosynthesis, cosmic microwave background, and large scale 
structure formation strongly constrain late time PTs.  At earlier stages of the 
cosmic evolution, thermal equilibrium erases most of the traces of PTs.  
Therefore, particle species which decouple early offer us perhaps the best probe 
PTs in the early universe.  Among candidate particles, possibly the most obvious is 
the TeV-scale dark matter (DM), which is expected to freeze out of thermal 
equilibrium around $O(10-100)$ GeV.  The successful prediction of 
its relic abundance, sometimes referred to as the WIMP miracle, is 
considered to be one the most important hints of new physics at the TeV-scale.  
Such a scenario is expected to be thoroughly probed at the LHC.

In this paper, we assess the sensitivity with which DM may probe the 
physics of PTs by exploring how PTs occurring nearly coincident with freeze 
out can modify the relic abundance calculation and alter the predicted relic 
density.  In PTs for which supercooling is non-negligible, we find that 
several competing effects contribute to an overall shift in the DM relic abundance 
(as compared to the usual calculation without a PT).  Two of these effects, the 
decoupling of non-relativistic species and the vacuum energy contribution to 
the Hubble expansion rate, tend to increase the relic abundance, while the 
entropy produced by the PT tends to decrease it.  The principal result of this 
paper is summarized in Eq.~(\ref{eq:finalfractionalchange}), and the central 
discussion will emphasize the role of vacuum energy during the PT 
\cite{Kolb:1979bt}, since that is the most novel aspect of this letter as compared 
to previous studies.  We find that a parametric tuning of order one percent can 
lead to an order unity dark matter abundance shift due to the presence of 
vacuum energy, assuming that a tuning of the cosmological constant  
sets the vacuum energy today.  In such situations, it may be possible to use 
DM as a probe of vacuum energy during the early universe by measuring the 
DM properties at terrestrial experiments and making mild assumptions about cosmology and UV completions of the effective field theory.  

This work is related to past papers which discuss moduli dilution,
such as \cite{Lyth:1995ka} (and hundreds of inflationary papers), in
that we calculate how the PT effects (including the vacuum energy)
alter the relic abundance.\footnote{It is also related to papers such
  as \cite{Kamionkowski:1990ni, Khlopov:1982, Khlopov:1983} and many others which consider the change
  in the relic density due to a change in the equation of state during
  the freeze out process.  Instead of listing all papers, interested
  readers can consider finding citations to and references within these
  papers.}  However, unlike the present work, most of these papers do
not consider the case of a PT which nearly coincides with freeze out,
nor do they consider the case of a low scale (e.g., electroweak scale)
PT with electroweak scale vacuum expectation values.
As in \cite{Cohen:2008nb,Feng:2009mn}, our calculation incorporates the
possibility that the dark matter annihilation cross section may change
after the freeze out, and as in \cite{Wainwright:2009mq}, we estimate
the dilution of dark matter due to a release of entropy at the PT.
However, we also include additional features of the PT such as the
changing vacuum energy.

This paper is organized as follows.  In Sec.~\ref{sec:genframe}, we
derive a generic parameterization with which one can discuss the effects of a 
PT on the DM relic abundance. In Sec.~\ref{sec:parametermapping}, we
use this parameterization to estimate the correction to the relic abundance in 
two toy models in which a real scalar field experiences a phase transition.  
In Sec.~\ref{sec:Conclusion} we summarize and briefly discuss which aspects 
of a generic model could be favorable for enhancing the effect of vacuum energy 
on the relic abundance.  Being a letter, we restrict ourselves to the highlights.  

Throughout the paper, we work in the FLRW spacetime $ds^2=dt^2 -a^2(t)
|d{\bf x}|^2$ in which $a(t)$ is a monotonically increasing function of $t$.  

\section{\label{sec:genframe}General framework}

In this section, we discuss the various ways in which phase
transitions can affect the relic density, and we provide a general
parameterization which is useful for analyzing specific models.

Integrating the thermally averaged Boltzmann equation, we obtain 
the number density of dark matter today ($a = a_0$ and $t = t_0$)
\begin{equation}
	n_{X}(t_{0})= x_0^{-3} \left(\int_{0}^{\ln x_0}\frac{d\ln(x)}{H} \sv \right)^{-1},
	\qquad x= \frac{a}{a_f}, 
	\quad x_0=\frac{a_0}{a_f} \label{eq:goodformula} 
\end{equation}
where $a_f$ corresponds to the scale factor at the time of the freeze out. 
We define the fractional deviation of the relic abundance as
\begin{align}\label{eq:fractional_dev}
	\delta n_X(t_0) = \frac{n_X(t_0)}{n_X^{(U)}(t_0)} - 1
\end{align}
where $n_X^{(U)}(t_0)$ is the ``usual'' relic density that one finds assuming 
that the PT does not occur.  In Eq.~(\ref{eq:goodformula}), the quantities that 
will be affected by the PT are the Hubble expansion rate $H(a)$, the thermally 
averaged cross section $\sv$ (also a function of $a$), and the dilution factor 
$x_0^{-3}=(a_f/a_0)^3 \ll 1$, which accounts for the expansion of the universe 
from freeze out until today (related to $T(a)$).

Suppose that a PT occurs after the time of the freeze out.  This PT can affect
$H(a)$ (though the energy density) in three ways: exotic energy, reheating, 
and decoupling.  First, the PT is a change in the vacuum state and is 
typically accompanied by a decrease in the vacuum energy.  For all 
(cosmological) intents and purposes, this vacuum energy behaves as a 
cosmological constant (CC).  Speaking more generally, we can collectively 
refer to the vacuum energy, cosmological constant, and any other non-thermal 
sources of energy (e.g., quintessence) as ``exotic energy.''  We assume that 
the exotic energy density can be written as 
$\rho(x) = \rho_{\ex} \kappa(x)$
with 
\be \label{eq:fitfunctionalform}
	\kappa(x)\approx
	\Theta ( (1+\delta)-x)
	+\Theta(x-(1+\delta)) 
	\left(
	1-\frac{\Delta\rho_{\ex}}{\rho_{\ex}} \right) \kappa_{2} (
	x
	 ) \, ,
\ee 
where $\Theta(z)$ is a step function and 
$\delta\equiv\frac{a_{PT}}{a_{f}}-1\lesssim1$ quantifies the delay between 
freeze out and the phase transition.  During the phase transition, the 
exotic energy decreases by $\Delta\rho_{\ex}>0 $, and the step function 
approximation corresponds to restricting ourselves to only phase transitions 
that occur on a time scale much shorter than $1/H$.  Such short time scale 
phase transitions are expected to be generic for models in which the thermal 
bounce action has a strong temperature dependence.\footnote{For a recent 
  discussion of situations with a longer time scale transitions, see for example
  \cite{Megevand:2007sv}.}  
In the case that the exotic energy is simply composed of vacuum energy plus 
a tuned cosmological constant,\footnote{This has been considered as an 
  acceptable possibility \cite{Linde:1974at,Ellis:1975ap}, and it is a consequence 
  of recently proposed string landscape scenario \cite{Polchinski:2006gy}.} 
we have $\Delta \rho_{\rm ex} \approx 0$ if the phase transition is of the second order or a smooth cross over and $\Delta \rho_{\rm ex} \approx \rho_{\rm ex}$ if the phase transition is first order with large supercooling.  
In the case $\Delta \rho_{\rm ex} \neq \rho_{\rm ex}$, the behavior of 
$\kappa_{2}(x)$ can parameterize quintessence dynamics which we assume 
decreases approximately as $\left(x \,  a_f / a_{PT} \right)^{-n_d}$ where
$n_d$ is a computable model dependent parameter.  We focus on phase
transitions that can be parameterized by a weakly coupled scalar field
description.

The remaining ways in which a PT can affect $H(a)$ are via the radiation 
energy density.  From energy conservation, the change in exotic energy 
$\Delta \rho_\ex$ must be compensated by a release of radiation energy, or 
equivalently, a reheating with entropy release $\Delta s$.  In addition, 
generically particle masses may depend upon the scalar field vacuum 
expectation value (VEV) and may increase during the phase transition 
(e.g., this is the case in the SM electroweak phase transition).  The heavier 
degrees of freedom can become non-relativistic and decouple.  Consequently, 
the remaining relativistic species have a relatively lower energy density.  We 
can parameterize this decoupling effect by writing the effective number of 
degrees of freedom for radiation energy $g_E$ and entropy $g_S$ as 
\begin{align}
	g_{E/S} \left( x \right) & = g_{E/S}(1) -
        h(x) \label{eq:gES_param} \\ 
	h (x) & = \frac{7}{8}N_{PT} \,
        \Theta(x-(1+\delta))+\frac{7}{8}N f (x) \label{eq:h_def}  \, ,
\end{align}
where ($N_{PT}$) $N$ represents the number of fermionic degrees of 
freedom which have (non-) adiabatically decoupled, and $f(x)$, which
rises from $0$ to $1$, is given by \eref{eq:fxeqapproxexpl}.

Treating all of the aforementioned effects as small perturbations and using $T(a)$
from \eref{eq:Tofa}, the modification to $H(a)$ can be expressed as
\begin{equation}\label{eq:H_linear}
	H\approx H_{R}^{(U)}(x)\left[1+\frac{\epsilon_{1}}{2}x^{4} \,
          \kappa(x)+\frac{2}{3}\epsilon_{2}
          \Theta(x-(1+\delta))+\frac{\epsilon_{31}\Theta(x-(1+\delta))+\epsilon_{32}f\left(x\right)}{6}\right]   
\end{equation}
where
\begin{equation}
	H_{R}^{(U)}(x)\equiv\frac{T_{f}^{2}}{3 \, M_{p} \,
          x^{2}}\sqrt{\frac{\pi^{2}}{10} \, g_{E}(T_{f})} 
\end{equation}
is the ``usual'' Hubble parameter in the absence of a PT, $T_f$ is the temperature 
at freeze out, and
\begin{subequations}\label{eq:epsiloni_def}
\begin{align}
	\epsilon_{1} & \equiv \frac{\rho_{\ex}}{\frac{\pi^2}{30} g_{E}(T_{f}) \, T_{f}^{4}}=\mbox{fractional energy of the exotic during freeze out} 
\label{eq:fractionalenergy} \\
	\epsilon_{2} & \equiv \left(1+3\, \delta \right)\frac{\Delta s}{\frac{2 \pi^2}{45} g_{S}(T_{f}) \, T_{f}^{3}}=\mbox{fractional entropy increase during PT}
\label{eq:fractionalentropyduringpt} \\
	\epsilon_{31} & \equiv\frac{\frac{7}{8}N_{PT}}{g_{E}(T_{f})}=\mbox{ fractional decoupling degrees of freedom during PT}
\label{eq:fractionaldecouplingduringPT} \\
\epsilon_{32}& \equiv\frac{\frac{7}{8}N}{g_{E}(T_{f})}=\mbox{ fractional decoupling degrees of freedom}
\label{eq:fractionaldecoupling}
\end{align}
\end{subequations}
are small, dimensionless quantities.  

Furthermore, if the dark matter is coupled to the scalar sector directly, a PT in the 
scalar sector may alter the annihilation cross section $\sv $.  This effect can be
parameterized by
\begin{equation}
	\sv =\sv^{(U)}\Bigl(1-\epsilon_{4}\, \Theta(x-(1+\delta))\Bigr) 
	\qquad \text{with} \qquad
	\epsilon_{4}\equiv-\frac{\Delta_{\sigma}}{\langle\sigma v\rangle^{(U)}} \, .
\label{eq:crosssecchange}\end{equation}
Since the derivation of Eq.~(\ref{eq:goodformula}) assumes that the dark matter 
is decoupled after $T_f$, we will assume that $\epsilon_{4}\gtrsim0$ in order 
to prevent re-thermalization due to an increase $\sv$. 

Finally, we turn our attention to the dilution factor $x_0^{-3}$.  Phase transitions 
occurring close to the freeze out can change the Hubble expansion rate, which 
in turn can cause dark matter to freeze out earlier or later.  We can parameterize 
this effect by approximating the freeze out temperature as 
\begin{equation}
	T_{f}\approx\frac{m_{X}}{\ln A}\left[1+\frac{\epsilon_{1}}{2}\left(\frac{1}{\ln A}+\ord{(\ln A)^{-2}\right)}\right]
\label{eq:freezeouttemp}\end{equation}
where 
\begin{equation}
	A\equiv\frac{N_{DM}3\sqrt{5}M_{p}\sqrt{m_{X}T_{f}}\langle\sigma v\rangle}{4\pi^{5/2}\sqrt{g_{E}(T_{f})}}\sim\exp[20] \, ,
\label{eq:approxexpressforA}\end{equation}
$m_X$ is the dark matter mass, and $N_{DM}$ counts the real dynamical 
degrees of freedom of the dark matter.  By also taking into account the 
effect of a late time entropy release associated with the PT, we obtain
\begin{equation}
	x_0 =  \left. \frac{a_0}{a_f} \right|_{\mathrm{usual}}  \times 
	\left[1+\frac{\epsilon_{1}}{2}\frac{1}{\ln A}+\frac{\epsilon_2}{3}\right]
\label{eq:a0overaffin}\end{equation}
where 
\begin{equation}\label{eq:a0_over_af_U}
	\left. \frac{a_{0}}{a_{f}} \right|_{\mathrm{usual}}\equiv\left(\frac{g_{S}(T_{f})}{g_{S}(T_{0})}\right)^{1/3}\frac{m_{X}}{T_{0}}\frac{1}{\ln A} 
\end{equation}
is the ``usual'' dilution factor in the absence of a PT, and $T_0$ is the temperature 
today.  In this paper we assume that the PT described by 
Eq.~(\ref{eq:fitfunctionalform}) is the last PT that generates appreciable entropy, 
but one can easily generalize \eref{eq:a0_over_af_U} to accommodate later 
PTs that generate more entropy.\footnote{In particular, we assume that QCD 
  phase transition is not a
  significant source of entropy \cite{Laermann:2003cv}.}

Putting everything together, we obtain a general parameterization of the 
changes to the dark matter relic abundance which are induced by a PT:
\begin{equation}\label{eq:finalfractionalchange}
	\delta n_X(t_0)  = c_1 \, \epsilon_1 + c_2 \, \epsilon_2 + c_{31} \, \epsilon_{31} + c_{32} \, \epsilon_{32} + c_4 \, \epsilon_4
\end{equation}
where the coefficients
\begin{subequations}
\begin{align}
	c_1 & \equiv \frac{1}{2}\left(\delta+\frac{\left(1+3\delta\right)}{n_d -3}\left(1-\frac{\Delta\rho_{\ex}}{\rho_{\ex}}\right)\right)-\frac{3}{2}\frac{1}{\ln A} \label{eq:c1_def}  \\
	c_2 & \equiv -\frac{1}{3}(1+2\delta) \label{eq:c2_def} \\
	c_{31} & \equiv \frac{1}{6}(1-\delta) \label{eq:c31_def} \\
	c_{32} & \equiv \frac{1}{6}\int_{1}^{a_{0}/a_{f}|_{\mathrm{usual}}}\frac{dx}{x^{2}}f\left(x\right) \label{eq:c32_def} \\
	c_4 & \equiv 1-\delta \label{eq:c4_def} 
\end{align}
\end{subequations}
are order one numbers that account for the delay between freeze out and the PT
(recall $\delta = a_{PT} / a_f -1 \gtrsim 0$).  Note that $c_{32}$ receives most of 
contributions from near the freeze out temperature.  The usefulness of this 
parameterization is that it is general enough to classify most phase transitions 
that can affect the DM relic abundance.  This is one of the main results of this 
paper.

\section{\label{sec:parametermapping} Phase transition effects as a function of Lagrangian parameters}

In the section, we discuss how the parameters of a scalar field theory map to
the freeze out modifying effects discussed in the previous section. In particular, 
we focus on a generic real scalar field for which the one-loop thermal effective 
potential is well-approximated by
\begin{align}\label{eq:GenSin_Veff}
	V_{\rm eff}(\phi,T) \approx \rhoex + \frac{1}{2} M^2 \phi^2 - \mathcal{E} \, \phi^3 + \frac{\lambda}{4} \phi^4 + c \, T^2 \phi^2  
\end{align}
where $M^2, \mathcal{E}, \lambda,$ and $c$ are free parameters.\footnote{ In
  order to treat $c$ as a free parameter, we must suppose that the
  $\phi$-sector is coupled to another sector, call it sector $X$,
  which is not strongly constrained phenomenologically.  The
  interaction between $\phi$ and sector $X$ can then be considered a
  nearly a free parameter and generates the thermal mass $c \, T^2$.  For
  instance, suppose a Yukawa coupling $\mathcal{L} \ni y \phi
  \bar{\psi} \psi$ where $\psi$ is a spin-1/2 $X$-sector field with
  $N$ dynamical degrees of freedom, and then $c \approx N y^2 / 48$.
  E.g., to obtain $c \approx 0.1$ one needs $y \approx 1.1$ if $N = 4$
  (Dirac fermion) and $y \approx 0.6$ if $N = 12$.  Moreover, the
  $X$-sector particles must be lighter than the PT temperature.
  Otherwise, Boltzmann suppression drives $c \to 0$.  } This can be viewed as 
the effective description of the dynamics of a large class of PTs with a tuned 
cosmological constant.  This simple description contains all the information that 
is necessary to discuss the vacuum energy contribution ($c_1 \epsilon_1$)
 and reheating contribution ($c_2 \epsilon_2$) 
to $\delta n_X(t_0)$.  The contributions from the decoupling 
($c_{31} \epsilon_{31}$ and $c_{32} \epsilon_{32} $) depend on additional 
details of the model and, as we will see, they have the dominant effect on the 
DM abundance.  Therefore, as far as we are concerned with 
the mapping of Lagrangian parameters to $c_i \epsilon_i$, we will focus 
our discussion on just $c_1 \epsilon_1$ and $c_2 \epsilon_2$.  Here, we also 
follow the traditional abuse of language in classifying the cosmological phase 
transitions as first order or second order dependent on whether or not
(transient) bubbles are involved during changes in the vacuum determining the 
1-particle state.

\subsection{$\mathcal{E} =0$, ``second order'' phase transition}
We first restrict our analysis to the case of $\mathcal{E}=0$. In this limit there is a
$\mathbb{Z}_2$ symmetry, and the finite temperature effective potential can be 
written as
\begin{align}\label{eq:Veff_1Dtoy}
	V_{\rm eff} (\phi,T) \approx \frac{\lambda}{4} \left( \phi^2 - v_\phi^2 \right)^2 + c \, T^2 \phi^2,
\end{align}
where $v_\phi = \sqrt{- M^2 /\lambda}$ is the VEV in the $\mathbb{Z}_2$ broken 
phase at $T=0$.  Because there is no cubic term, no sub-horizon bubbles are 
involved as the vacuum changes from $\phi=0$ to $\phi=v_\phi$ at the PT.\footnote{Horizon sized domain walls do form,
  however \cite{Kibble:1976sj}.}  
The temperature at the beginning of the PT can be approximately mapped to the
Lagrangian parameters as $(T_{PT}^-)^2 =\frac{ \lambda}{2c} v_{\phi}^2 $.
By requiring that the exotic energy be zero today when $T=0$, we find the 
exotic energy at the time of the phase transition to be 
\begin{align}
	\rho_{\ex} = V_{\rm eff}(0,0) = \frac{\lambda}{4} v_{\phi}^4 = \frac{c^2}{\lambda} (T_{PT}^-)^4. \label{eq:vacenergyzeroT}
\end{align}
Therefore using Eqs.~(\ref{eq:fractionalenergy}) and (\ref{eq:c1_def}) the exotic 
energy contribution is given by (for $\delta < 1$)
\begin{align}\label{eq:cc_2ndorder}
	c_1 \epsilon_1 \approx  \frac{\delta}{2 g_E} \frac{c^2}{\lambda}  \sim \frac{1}{10} \frac{1}{g_E} c^2 \frac{v_{\phi}^2}{m_\phi^2} 
\end{align}
where $m_\phi^2 = 2 \lambda v_{\phi}^2$ is the approximate scalar mass in the 
$\phi = v_{\phi}$ vacuum, and typically $g_E \gtrsim 100$.  

In the minimal scenario of the SM supplemented by a DM sector, one finds 
$c^2_{\rm SM}/\lambda_{\rm SM} \approx 0.28$ where $c_{\rm SM}$ is dominated
by the top Yukawa and does not take into account the coupling of DM to the 
Higgs sector.  If electroweak symmetry breaking occurs soon after the dark 
matter freeze out, Eq.~(\ref{eq:cc_2ndorder}) 
allows us to estimate that the DM relic abundance will experience a fractional 
change at the order of $10^{-3}$ due to each of the CC 
effect.
Moreover, soon after the electroweak phase transition, the heavy quarks 
decouple and $N \sim 20$ fermionic degrees of freedom are lost from the tally 
of relativistic species.  Consequently, the ratio 
\begin{align}
	\frac{c_1 \epsilon_1}{ c_{32} \epsilon_{32}} \sim -\frac{c^2}{\lambda} \frac{1}{N} \lesssim 1
\end{align}
is small, and we expect that the shift in the relic abundance is dominated by the 
decoupling of these heavy degrees of freedom.

In the SM, the exotic energy effect is subdominant, but Eq.~(\ref{eq:cc_2ndorder}) 
provides a guide to constructing models with enhanced $c_1 \epsilon_1$.  This 
term can be made larger if $v_\phi^2 /m_\phi^2 \gg 1$, which could be realized 
by invoking fine tuning or some additional symmetry to generate a flat potential.  
Alternatively, one could contrive a model in which 
$\rhoex \gg T_{f}^4 \geq (T_{PT}^-)^4$ and thereby enhance $\epsilon_1$ directly.  
Such a scenario can be naturally realized if supercooling occurs, as in the case 
of a ``first order'' PT.  We now turn our attention to this scenario.  

\subsection{$\mathcal{E}\neq0$, supercooling and ``first order'' phase transition}

At $T=0$, the general potential in \eref{eq:GenSin_Veff} has extrema at 
\begin{align}
	\phi=0 \quad \text{and} \quad \phi= v_{\phi} = \frac{3 \mathcal{E}}{2 \lambda} \left( 1 + \sqrt{1 - \frac{8}{9} \alpha_0} \right) \, ,
\end{align}
where we have introduced the dimensionless quantity 
$\alpha_0 \equiv \lambda M^2 / 2 \mathcal{E}^2$, which controls the vacuum 
structure.  For $\alpha_0 > 1$, $\phi = 0$ is the true vacuum; for 
$0 < \alpha_0 < 1$, $\phi = v_{\phi}$ is the true vacuum while $\phi = 0$ is 
metastable; and for $\alpha_0 < 0$, $\phi = 0$ becomes unstable.  The barrier 
separating the metastable and true vacua has a height (for $0<\alpha_0<1$)
\begin{align}\label{eq:GenSing_Vbar}
	V_{\mathrm{barrier}} = \frac{4 \mathcal{E}^4 \alpha_0^3}{27 \lambda^3} \Bigl( 1 + \ord{\alpha_0} \Bigr)
\end{align}
which vanishes rapidly as $\alpha_0 \to 0$.  As in \eref{eq:vacenergyzeroT}, by 
requiring the exotic energy to vanish today, we calculate the exotic energy prior 
to the PT to be 
\begin{align}\label{eq:GenSing_rhoex}
	\rho_{\rm ex} = \frac{\mathcal{E}^4}{8 \lambda^3} \left[ 27 - 36 \, \alpha_0 + 8 \, \alpha_0^2 + 27 \left(1 - \frac{8}{9} \alpha_0 \right)^{3/2} \right]  
\end{align}
and note that all of this energy is converted into radiation at the phase 
transition (i.e., $\Delta \rho_{\rm ex} = \rho_{\rm ex}$).  

In order to compute the CC's effect on the relic abundance, we need to
know the PT temperature $T_{PT}^-$, or equivalently the amount of 
supercooling, which has an interesting dependence on $\alpha_0$.
We require $\alpha_0 < 1$ such that there exists a temperature 
\begin{align}\label{eq:Tc}
	T_c = \mathcal{E} \sqrt{\frac{1 - \alpha_0}{\lambda \, c}}
\end{align}
below which the symmetric phase $\phi=0$ becomes metastable.  The PT 
begins at a temperature $T_{PT}^- < T_c$ when the bubble nucleation rate per 
Hubble volume $\Gamma \, H^{-3} \sim T^4 e^{-S^{(3)}/T} \, H^{-3}$ is comparable 
to Hubble expansion rate $H \sim T^2/M_p$.  Here $S^{(3)}$ is the action of the 
O(3) symmetric bounce.  For an electroweak scale phase transition this condition 
is satisfied when $S^{(3)} / T$ drops below approximately $140$ 
\cite{Quiros:1999jp}.  Provided that the potential can be expressed in the form 
of \eref{eq:GenSin_Veff}, then the action is well-approximated by the empirical 
formula \cite{Dine:1992wr}
\begin{align}
	\frac{S^{(3)}}{T} &\approx 13.7 \, \frac{\mathcal{E}}{T} \left(\frac{\alpha}{\lambda} \right)^{3/2} f(\alpha) \label{eq:S3overT}  \\
	f(\alpha) &\equiv 1 + \frac{\alpha}{4} \left( 1 + \frac{2.4}{1-\alpha} + \frac{0.26}{(1-\alpha)^2} \right)
\end{align}
where the temperature dependence is parameterized by 
$\alpha(T) = \alpha_0( 1 - T^2 / T_0^2)$, and $T_0^2=  -M^2 / (2 c)$ can be 
positive or negative.

The PT temperature is constrained by $\mathrm{Max \, }[T_0^2, 0] <
(T_{PT}^-)^2 < T_c^2$ where the lower bound depends on the sign of
$\alpha_0$.  We will discuss the two cases separately.  For $\alpha_0
> 0$ (or $T_0^2 < 0$), the vacuum $\phi = 0$ remains metastable as $T
\to 0$.  This suggests that the PT temperature can be arbitrarily low,
and in this limit of large supercooling the CC effect may be
arbitrarily large.  Unfortunately, if the barrier persists as $T \to
0$, it is possible that the PT does not occur at any
temperature -- a obviously unphysical scenario in the case of the electroweak 
phase transition.  This follows from
the observation that for $\alpha_0 > 0$, $S^{(3)} / T$ has a minimum
at $T\neq 0$: at low temperatures $S^{(3)} / T$ grows due to the
explicit factor of $T$ in the denominator, and at high temperatures
$f(\alpha)$ diverges as $\alpha \to 1$.  Over some of the parameter space, 
the inequality $S^{(3)} / T \lesssim 140$ is not satisfied at any
temperature, and the PT does not occur.  In particular, if $\alpha_0$ is 
close to one, then $\alpha > \alpha_0 \approx 1$ 
at all temperatures, and it is very difficult for the PT to proceed.  
Therefore, if we require that the PT must occur via
thermal bubble nucleation, we obtain an upper bound on $\alpha_0$.
For the case $\alpha_0 < 0$, the PT necessarily occurs
at a temperature $T_{PT}^- > T_0 > 0$, since the $\phi = 0$ vacuum
becomes perturbatively unstable below $T_0$.  This case has the
drawback that supercooling cannot last an arbitrarily long time, but
on the other hand, one is guaranteed that the PT
proceeds.

\begin{figure}[t]
\begin{center}
\includegraphics[width=0.6\textwidth]{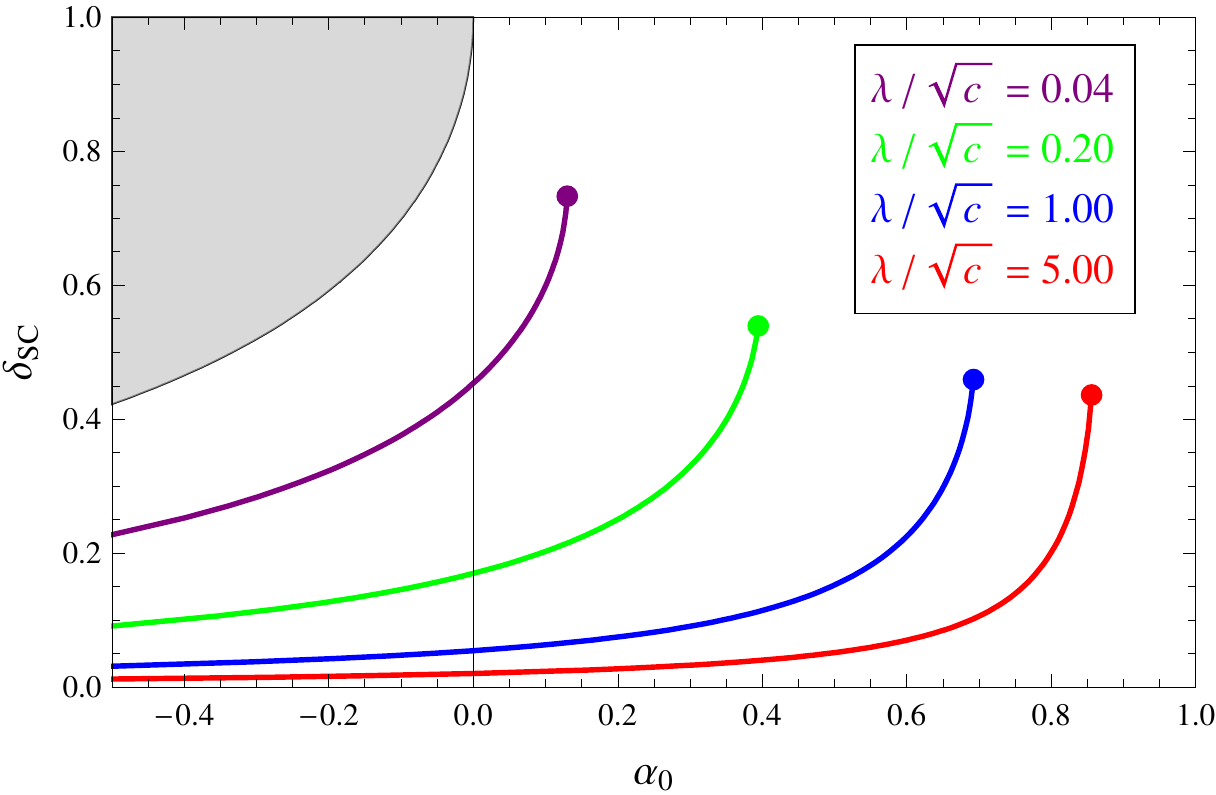} 
\caption{\label{fig:deltaSC} We have plotted the amount by which the
  phase transition temperature drops below the critical temperature,
  quantified by $\delta_{\rm SC}$, against the parameter $\alpha_0$ which
  controls the height of the barrier.  These curves only depend on the
  parametric combination $\lambda / \sqrt{c}$.  The amount of
  supercooling grows as $\alpha_0$ is made larger, but reaches a
  finite maximum $ \delta_{\rm SC}^{(max)} \lesssim \ord{1}$ at a value of
  $\alpha_0$ that depends on the ratio $\lambda / \sqrt{c}$.  }
\end{center}
\end{figure}

Provided that the PT does occur, we define 
\begin{align}\label{eq:GenSing_dSC}
\delta_{\rm SC} = 1 - \frac{T_{PT}^{-}}{T_c}
\end{align}
which ranges from $0$ to $1$ and quantifies the amount of supercooling.  
Using $\delta_{\rm SC}$ to parameterize the temperature dependence, we can 
rewrite \eref{eq:S3overT} in the form
\begin{align}
	\left. \frac{S^{(3)}}{T} \right|_{T_{PT}^-} \approx 
	\left( \frac{\lambda}{\sqrt{c}} \right)^{-1} \frac{1}{\sqrt{1-\alpha_0}} 
	\left[ \frac{a_{-2}}{\delta_{SC}^2} + \frac{a_{-1}}{\delta_{SC}} + a_0 + a_1 \delta_{SC} + \ord{\delta_{SC}^2} \right] \, ,  
\end{align}
where the $a_i$ are functions of $\alpha_0$.  We require 
$\left. S^{(3)}/T \right|_{T_{PT}^-} = 140$ and solve for $\delta_{\rm SC}$, which 
we have plotted in Figure \ref{fig:deltaSC}.  The supercooling grows with 
increasing $\alpha_0$ and decreasing $\lambda / \sqrt{c}$ as the barrier and 
bounce action are made larger.  In the shaded region the lower bound on 
$T_{PT}^- > T_0$ is not satisfied.  The amount of supercooling is typically 
of the order $\delta_{\rm SC}\lesssim 0.5$ which implies $T_{PT}^- \gtrsim T_c / 2$.  
Above a finite value of $\alpha_0$ (indicated by a dot) the barrier becomes 
insurmountably large, and the universe becomes trapped in the metastable 
vacuum.  The existence of this upper bound on $\alpha_0$ does not allow a 
phenomenologically viable, arbitrarily large supercooling, contrary to naive 
expectations.  The largest amount of supercooling is achieved for 
$\lambda / \sqrt{c} \ll 1$ and $\alpha_0 \gtrsim 0$.  In this parameter regime the 
exotic energy is large (see \eref{eq:GenSing_rhoex}), and the metastable vacuum 
is separated from the true vacuum by a small barrier (see \eref{eq:GenSing_Vbar}).

\begin{figure}[t]
\begin{center}
\includegraphics[width=0.45\textwidth]{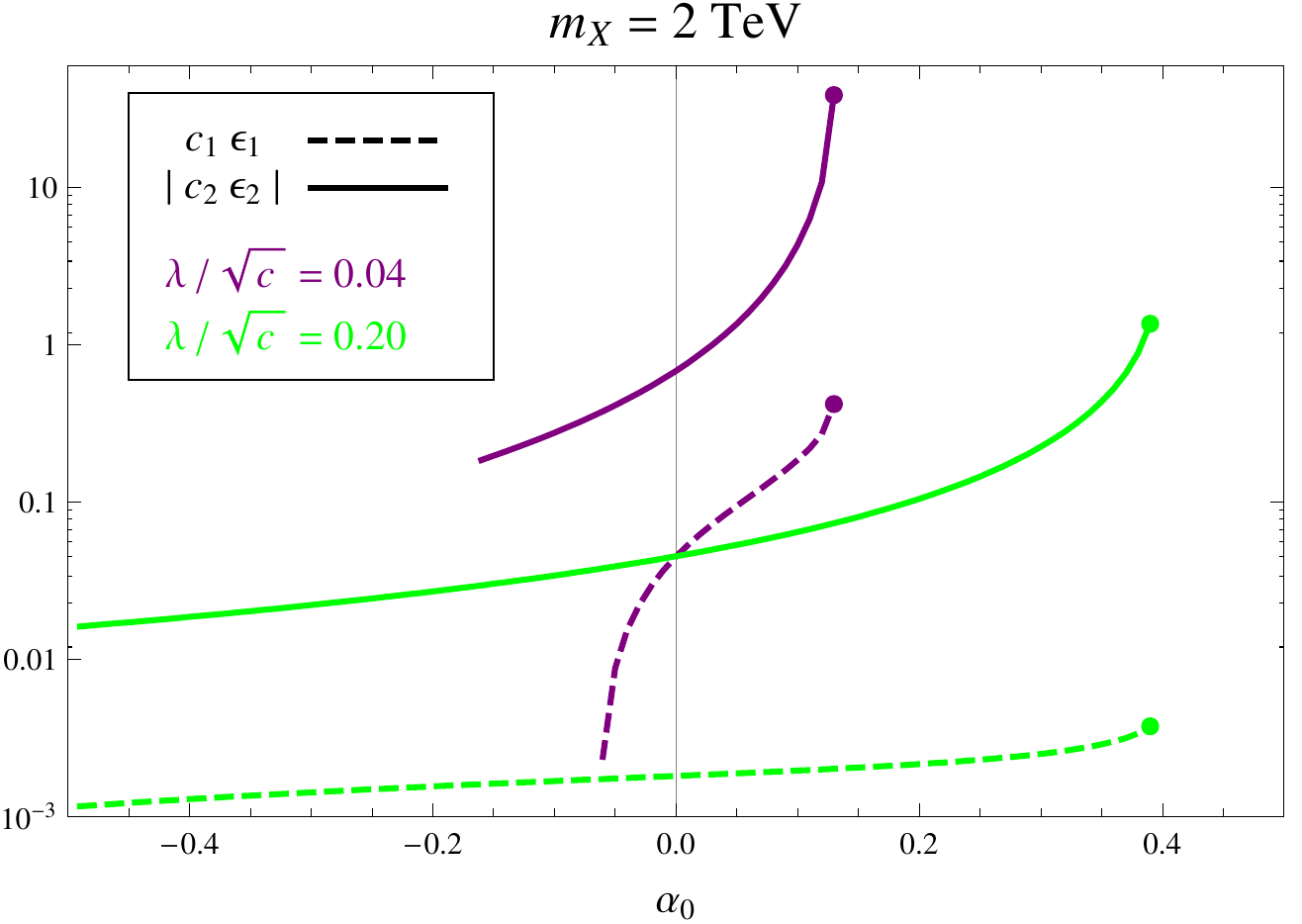} \hfill
\includegraphics[width=0.45\textwidth]{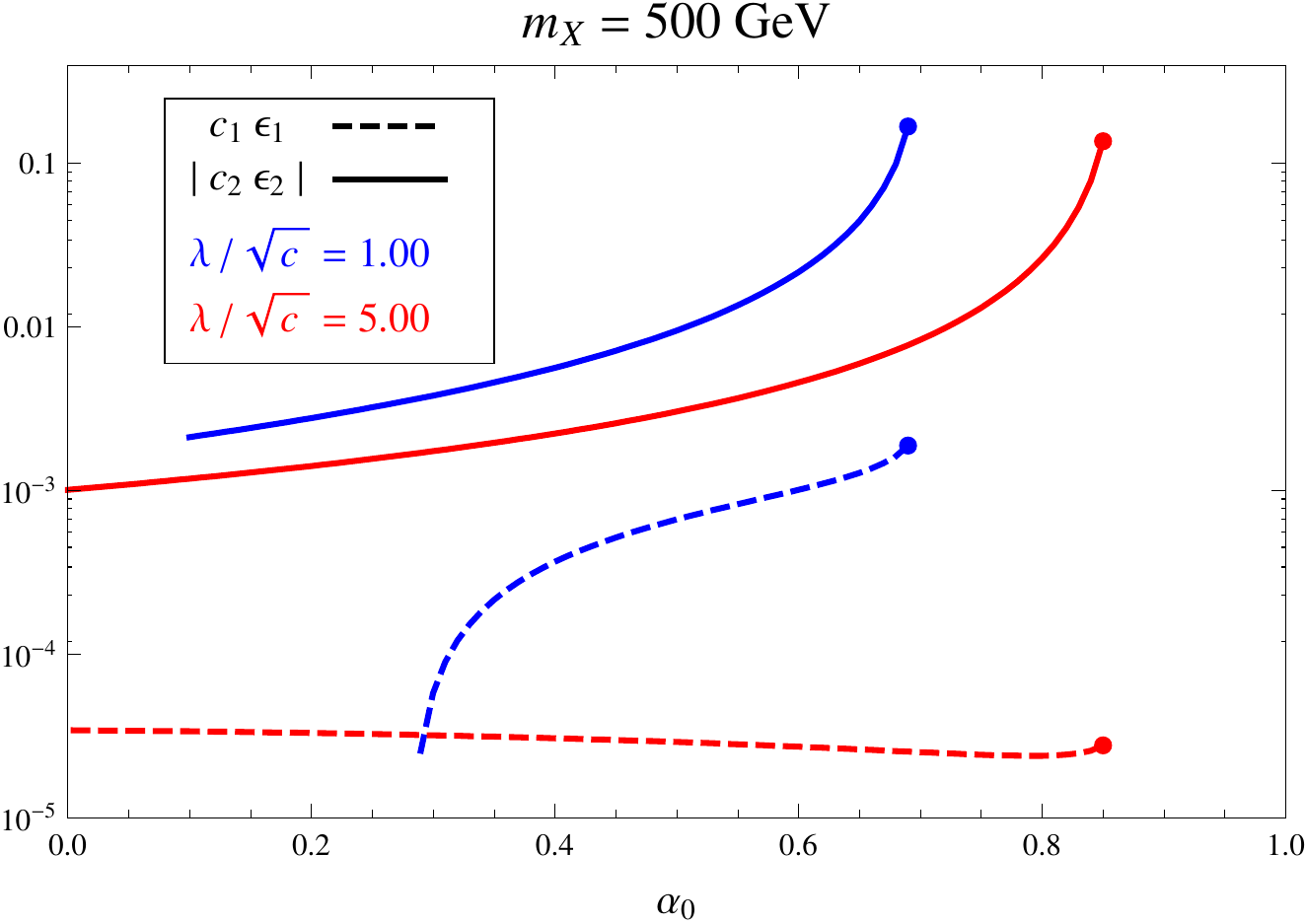}
\caption{\label{fig:c1e1_and_c2e2} The fractional shift in the dark
  matter relic abundance due to the exotic energy (dashed, $c_1
  \epsilon_1$) and the reheating (solid, $c_2 \epsilon_2$).  Note that
  the two figures have different scales, and that we have plotted
  $|c_2 \epsilon_2|$ since this quantity is negative.  When
  $\lambda/\sqrt{c}$ is smaller than 0.04, then one may enter a regime
  of large supercooling for tuned values of $\alpha_0$.  The reheating
  effect dominates by an order of magnitude or more.  The contours
  extend over a finite range of $\alpha_0$ because for larger
  $\alpha_0$ the PT does not occur, and for smaller $\alpha_0$ the PT
  occurs before freeze out.  
  Since our analytical approximation breaks down when $c_i \epsilon_i \sim O(1)$, the extrapolation into this region should only be treated as an indication of possible size of the effect.  }
\end{center}
\end{figure}

We have calculated the exotic energy and reheating contributions to the relic
abundance shift by using \eref{eq:finalfractionalchange}, and we present the 
results in Figure \ref{fig:c1e1_and_c2e2}.  In generating these plots, we have 
fixed $c=0.1$, $\mathcal{E} = 5$ GeV, and $g_{E/S} = 106.75$ (SM degrees of 
freedom\footnote{We choose this value as a fiducial reference.  Realistically, for 
  these parameters the PT occurs at $T_{PT}^- \approx 1 - 100$ GeV, which could be 
  later than the electroweak phase transition.  In that case, some of the SM degrees 
  of freedom would have decoupled, $g_{E/S}$ would be smaller, and the 
  $\epsilon_i$ would be relatively larger.}) 
while allowing $\alpha_0$ to vary.  We select two values for the dark matter mass, 
which in turn fixes the freeze out temperature via \eref{eq:freezeouttemp}.  For 
the heavier case $m_X = 2$ TeV, the freeze out occurs quite early, and if 
$\lambda / \sqrt{c} = 1.00,5.00$ (which are not shown) the PT would occur much 
later, in the limit where our analytic approximations break down (i.e., $\delta > 1$).  
Some of the curves are truncated at small $\alpha_0$, because we require that 
the PT occur after the freeze out (i.e., $\delta > 0$), and the phase transition 
temperature increases with decreasing $\alpha_0$ (see Figure \ref{fig:deltaSC}).  
It is also for this reason that, the $\lambda / \sqrt{c} = 0.04$ and $0.2$ curves are 
entirely absent from the $m_X = 500$ GeV plot.  

These figures indicate that the exotic energy effect on the relic abundance is 
typically on the order of $10^{-3}$ and is subdominant to the reheating effect by 
an order of magnitude.  Both contributions become larger in the limit of large 
supercooling where $\lambda /\sqrt{c}$ is small and $\alpha_0$ approaches its 
maximal value.  For smaller values of $\lambda/\sqrt{c}$ a brief period of inflation 
might even be possible.
The curves 
$\left\{ \lambda / \sqrt{c} = 0.04, m_X = 2 \text{ TeV} \right\}$ and 
$\left\{ \lambda / \sqrt{c} = 1.00, m_X = 500 \GeV \right\}$ illustrate the parametric 
tuning of $\alpha_0$ that is required to achieve a large correction to the relic 
abundance.  If $\alpha_0$ is made too large, the PT does not occur, and if 
$\alpha_0$ is made too small, the PT occurs before freeze out.  Comparing the 
$m_X = 2$ TeV and $m_X = 500$ GeV plots reveals the parametric tuning that 
must occur between the DM and scalar sectors.  If the DM mass is small, for 
example, then the parameters of the scalar sector must conspire to generate a 
low scale PT, otherwise the PT occurs too early and decouples from the physics
of the freeze out.

\section{Conclusion}\label{sec:Conclusion}
If the properties of dark matter can be measured accurately in
laboratories, the information that these experiments yield can be used
to probe the properties of early universe phase transitions.
This is a particularly exciting prospect
given that phase transition physics incorporates the energy densities
of the false vacuum and the cosmological constant, and thereby it
provides an empirical method to directly probe the tuning of the
cosmological constant.  With this in mind, we have developed a general
parameterization to characterize the effects of a single field phase
transition on the thermal dark matter relic abundance in a freeze out
scenario.

In the context of the SM (supplemented by a DM candidate) and assuming
a tuned cosmological constant, we find that the exotic energy
(i.e.~the Higgs field vacuum energy plus the cosmological constant
energy) leads to a fractional increase in the dark matter abundance by
$O(10^{-3})$.  The dominant change in the dark matter
abundance comes from a decoupling of relativistic degrees of freedom
near the time of the freeze out, which leads to a fractional increase
in the relic abundance of order $10^{-2}$.  
Without extreme tuning,
we expect that most second order PTs share the characteristics of the
SM case.

In the case of a second order phase transition, models with a very flat potential 
(i.e., $m_{\phi}^2 \lesssim H_{PT}$) generally give a large dark matter abundance 
shift via the exotic energy contribution.  In this limit, Hubble friction can enhance 
the supercooling as in the case of slow-roll inflation (as signaled by the 
enhancement attendant with large $v_\phi/m_\phi$ in Eq.~(\ref{eq:cc_2ndorder})).  
Although pseudo-Nambu-Goldstone boson models may be useful for producing 
such flat potentials, the required hierarchies can be somewhat unnatural during 
the electroweak phase transition since $H_{PT} \sim 10^{-14}$ GeV.  

In first order phase transitions with supercooling, there is a somewhat surprising 
theoretical upper limit on the duration of supercooling which follows from the fact 
that the bubble nucleation rate is not a monotonically decreasing function of time.  
In certain parametric regimes, the phase transition never occurs.  Close to this 
failed phase transition case, the maximum fractional increase in the relic 
abundance due to the exotic energy effect can become $O(0.1)$ and 
due to the reheating effect can become $O(1)$.  However, reaching 
these large magnitudes requires some degree of parametric tuning.  As the 
parameters deviate from their tuned values, either the PT will not occur at all, or 
it will occur before the freeze out.

In order for dark matter freeze out to act as a probe of the phase transition, as we 
have considered, it must be the case that freeze out occurs soon before or 
concurrently with the phase transition.  Since phase transitions typically occur 
at electroweak scale temperatures or higher and since the mass of weakly 
interacting dark matter is typically 20 times larger than the freeze out temperature, 
these DM particles must be heavy, and they may be difficult to discover at the LHC.  

It is nonetheless an exciting prospect that LHC and other experiments sensitive 
to dark matter's non-gravitational interaction properties may unveil a new probe 
of dark energy.  This is particularly interesting given that there is almost no other 
way to probe the conjecture of a tuned cosmological constant.\footnote{There are
  generic theoretical limitations on empirical reconstruction of the
  phase transition scenario.  This study will be presented elsewhere
  \cite{clwinprep2011-1}.}

\section{Acknowledgements}
We thank Lisa Everett and Sean Tulin for useful correspondence.  DJHC
and AJL were supported in part by the DOE through grant
DE-FG02-95ER40896.  LTW is supported by the NSF under grant
PHY-0756966 and the DOE Early Career Award under grant DE-SC0003930.

\appendix

\section{Derivation of PT induced change in the degree of freedom\label{sub:Derivation-of-PT-induced-dof-change}}

We begin with the well-known formula for the energy density of a gas of 
fermions at temperature $T$ with $N$ dynamical degrees of freedom:
\begin{align}
	\rho(T) = N \int \frac{d^3 p}{(2 \pi)^3} \frac{E_p}{1 + e^{E_p/T}} \, .
\end{align}
The gas has an effective number of degrees of freedom $g_E$ given implicitly 
by $ \rho(T) =\frac{\pi^{2}}{30} \, g_{E}(T) \, T^{4}$.  We can parameterize 
the decrease in $g_E$ due to the decoupling of the fermionic gas by writing 
\begin{align}
	g_E(T) = g_E(T_f) - \frac{7}{8} \, N f \left( a/a_f \right)
\end{align}
where 
\begin{equation}
	f\left(x=a/a_f\right)=\left(\frac{7}{8} \frac{\pi^2}{30}\right)^{-1}\int\frac{d^{3}p}{(2\pi)^{3}}E_{p}\left[\frac{1}{T_{f}^{4}}\frac{1}{e^{\frac{E_{p}}{T_{f}}}+1}-\frac{1}{T^4(a_f x)}\frac{1}{e^{\frac{E_{p}}{T(a_f x)}}+1}\right]\label{eq:fxeq} \, .
\end{equation}
The temperature $T = T(a)$ is given by \eref{eq:Tofa} to leading order in the perturbations $\epsilon_i$.  Since $f$ already multiplies a small term in \eref{eq:h_def}, we need only keep the leading factor in \eref{eq:Tofa} which is $T = T_f \, a_f / a = T_f / x$.  This lets us write \eref{eq:fxeq} as 
\begin{equation}
	f\left(x\right)=\frac{8}{7}\left(\frac{30}{\pi^{2}}\right)\int\frac{d^{3}p}{(2\pi)^{3}}\frac{E_{p}}{T_{f}^{4}}\left[\frac{1}{e^{\frac{E_{p}}{T_{f}}}+1}-\frac{x^{4}}{e^{\frac{xE_{p}}{T_{f}}}+1}\right]\label{eq:fxeqapproxexpl} \, .
\end{equation}
Note that $f(x)$ increases from $f(1)=0$ to $f(\infty) \approx 1$.  Due to the 
exponential temperature dependence, the transition to $f \approx 1$ occurs at 
$T\approx m_{N}$ and is smoothly steplike over a time scale $\Delta t \approx 1/H$. 
In this discussion we have assumed $E_p = \sqrt{ {\bf p}^2 + m_N^2}$ with $m_N$ 
constant, that is, we neglect any change in the mass of the particle as a function of 
time.  This assumption is valid sufficiently far after the PT such that the scalar VEV 
and field-dependent masses have approximately stopped varying.

\section{Derivation of $T_{PT}^+$, $\Delta s$, and $T(a)$}\label{sub:deriveTofa}

In this appendix, we calculate the temperature after the phase transition 
$T_{PT}^+$ by imposing energy conservation at the PT.  This allows us to 
calculate $\Delta s$ and $\epsilon_2$ in terms of $\drhoex$.  Provided that there 
is a negligible change in $a \approx a_{PT}$ during reheating, energy conservation 
can be written as
\begin{align}\label{eq:EnergyConsv}
	\frac{\pi^2}{30} \, g_E ( T_{PT}^-
	) \left( T_{PT}^- \right)^4 + \drhoex &= \frac{\pi^2}{30} \, g_E ( T_{PT}^+
	) \left( T_{PT}^+ \right)^4 \, .
\end{align}
Using the perturbative expansions introduced in Section \ref{sec:genframe}, 
\eref{eq:EnergyConsv} can be solved for $T_{PT}^+$ at leading order to obtain
\begin{align}\label{eq:TPTplus}
	T_{PT}^+ & \approx T_{PT}^- \left[1
	+ \frac{1}{4} \, \epsilon_{31} 
	+ \frac{1}{4} \frac{\drhoex }{\frac{\pi^2}{30} \, g_E(T_f) \, (T_{PT}^-)^4}
	\right]
\end{align}
where $\epsilon_{31}$ is given by \eref{eq:fractionaldecouplingduringPT}.  As 
expected, the amount of exotic energy released $\drhoex > 0$ controls the 
reheating from $T_{PT}^-$ to $T_{PT}^+$.  Additionally, the reheating is larger 
when more species non-adiabatically decouple (larger $\epsilon_{31}$), because 
there are fewer degrees of freedom after the PT to distribute $\Delta \rho_{\rm ex}$ 
over, which makes them comparatively hotter.  

Similarly, we can calculate the entropy density increase at the PT.  Writing 
the entropy density as $s(T) = \frac{2 \pi^2}{45} g_S(T) T^3$, we can calculate 
$\Delta s$ as 
\begin{align}
	\Delta s 
	& =\frac{2\pi^{2}}{45} \Bigl\{g_{S}(T_{PT}^{+}
	)(T_{PT}^{+})^{3}-g_{S}(T_{PT}^{-}
	)(T_{PT}^{-})^{3}\Bigr\}   \\
	& \approx \frac{2 \pi^2}{45} \left\{
	-\frac{g_E(T_f)}{g_S(T_f)} \epsilon_{31}
	+ \frac{3}{4} \left[ 
	\epsilon_{31} + \frac{\drhoex}{\frac{\pi^2}{30} g_E(T_f) (T_{PT}^-)^4}
	\right]
	\right\} g_S(T_f) \left( T_{PT}^- \right)^3 
\label{eq:relationshipbtwdeltasandrhoexotic}
\end{align}
where we have used \eref{eq:TPTplus} and linearized in perturbations.  We can 
calculate $\epsilon_2$, given by \eref{eq:fractionalentropyduringpt}, by noting 
$T_{PT}^- \, a_{PT} \approx T_f \, a_f$ and $g_S(T_f) \approx g_E(T_f)$ up to 
higher order terms.  Doing so yields
\begin{align}\label{eq:epsilon2_intermsof_L}
	\epsilon_2 & \approx- \frac{1}{4} \, \epsilon_{31} + \frac{3}{4} \frac{\drhoex}{\frac{\pi^2}{30} g_E(T_f) (T_{PT}^-)^4} \, .
\end{align}
These expressions for $\Delta s$ and $\epsilon_2$ illustrate that the entropy 
increase at the PT is controlled by the amount of latent heat released and the 
number of particles that non-adiabatically decouple.

Lastly, we will solve the equation of entropy conservation for $T(a)$.  The entropy 
per comoving volume $S = s \, a^3$ is conserved except for the entropy injection 
at reheating, which is assumed to occur rapidly at $a_{PT}$.  Entropy conservation 
may be expressed as
\begin{equation}\label{eq:EntropyConsv}
	g_{S}(T
	) \, T^{3}a^{3}=g_{S}(T_{f}) \, T_{f}^{3}a_{f}^{3}+\Theta(a-a_{PT}) \, a_{PT}^{3} \, \left( \frac{2 \pi^2}{45} \right)^{-1} \Delta s
\end{equation}
and implicitly defines $T(a)$.  To solve for $T$ we use \eref{eq:gES_param} to 
expand $g_S(T)$ then linearize in $h$ and $\Delta s$ to obtain 
\begin{align}
	T(a) \approx T_f \frac{a_f}{a} \left[ 1 
	+ \frac{1}{3} \frac{h(a/a_f)}{g_S(T_f)} 
	+ \Theta \left( a - a_{PT} \right) \frac{1}{3} \left( \frac{a_{PT}}{a_f} \right)^3 \frac{\Delta s}{\frac{2 \pi^2}{45} g_S(T_f) \, T_f^3} \right] \, .
\end{align}
Further expanding $h$ using \eref{eq:h_def}, approximating 
$g_S(T_f) \approx g_E(T_f)$, and applying \eref{eq:fractionalentropyduringpt} 
we obtain the final expression,
\begin{align}\label{eq:Tofa}
	T(a) \approx T_f \frac{a_f}{a} \left[ 1 
	+ \frac{1}{3} \epsilon_{32} \, f(a / a_f)
	+ \Theta \left( a - a_{PT} \right) \frac{1}{3} \left( 
	 \epsilon_{31} 
	+ \epsilon_2
	\right)
	\right] \, .
\end{align}
After the PT, the exotic energy component behaves approximately adiabatically.

\bibliographystyle{jhep}
\bibliography{references}

\end{document}